\documentclass{article}
\usepackage{spconf}

\usepackage[english]{babel}
\usepackage[utf8x]{inputenc}
\usepackage[T1]{fontenc}

\usepackage{amsmath,amssymb,mathrsfs,bbm}
\usepackage{graphicx}
\usepackage[colorinlistoftodos]{todonotes}
\usepackage{multirow}
\usepackage{makecell}

\usepackage{psfrag,epsfig,graphics}
\usepackage{amsmath,amsthm,amssymb,multirow}
\usepackage{mathbbol}
\usepackage{amssymb}   
\usepackage{url}  
\usepackage{xstring}
\usepackage{cite}

\usepackage{mathabx} %

\DeclareSymbolFontAlphabet{\amsmathbb}{AMSb}%

\newcommand{\cp}[1]{\ifmmode {\mathcal{#1}}\else ${\mathcal{#1}}$\fi}
	
\newcommand{\bA}{\boldsymbol{A}}
\newcommand{\bB}{\boldsymbol{B}}
\newcommand{\bC}{\boldsymbol{C}}

\newcommand{\bI}{\boldsymbol{I}}

\newcommand{\bS}{\boldsymbol{S}}

\newcommand{\bT}{\boldsymbol{T}}

\newcommand{\bV}{\boldsymbol{V}}

\newcommand{\bX}{\boldsymbol{X}}

\newcommand{\ba}{\boldsymbol{a}}
\newcommand{\bb}{\boldsymbol{b}}
\newcommand{\bc}{\boldsymbol{c}}

\newcommand{\bx}{\boldsymbol{x}}

\newcommand\tensor[1]{%
  \ifcat\noexpand#1\relax %
    \mathbb{#1}%
  \else
      \if\relax\detokenize\expandafter{\romannumeral-0#1}\relax  %
        \mathbb{#1}
      \else
        \boldsymbol{\mathcal{#1}}%
      \fi
  \fi }

\usepackage{color}  %
\def\cred{\textcolor{red}}

\definecolor{darkgreen}{rgb}{0., 0.4, 0.}

\usepackage[lined,linesnumbered,ruled]{algorithm2e}

\renewcommand{\cred}{}

\makeatletter
\renewcommand{\paragraph}{%
  \@startsection{paragraph}{4}%
  {\z@}{0.1ex \@plus 1ex \@minus .2ex}{-1em}%
  {\normalfont\normalsize\bfseries}%
}
\makeatother

\title{Coupled CP tensor decomposition with shared and distinct \\ components for multi-task fMRI data fusion\\\vspace{-0.05cm}}

\name{R. A. Borsoi$^1$, I. Lehmann$^2$, M. A. B. S. Akhonda$^3$, V. D. Calhoun$^4$, K. Usevich$^1$, D. Brie$^1$, T. Adali$^3$ \vspace{-0.15cm}
\thanks{This work was supported in part by NSF-NCS 1631838, NSF 2112455, and NIH grants R01 MH118695, R01 MH123610, and R01 AG073949, and in part by the German Research Foundation under grant SCHR 1384/3-2. T.~Adali's visit to CRAN was supported in part by the Universit\'e de Lorraine.} }%

\address{
\small $^1$ Universit\'e de Lorraine, CNRS, CRAN, Vandoeuvre-l\`es-Nancy, France \\[-0.05cm]
\small $^2$ Signal and System Theory Group, Paderborn University, Paderborn, Germany \\[-0.05cm]
\small $^3$ Dept. of CSEE, University of Maryland Baltimore County, Baltimore, MD, USA \\[-0.05cm]
\small  $^4$ Tri-institutional Center for Translational Research in Neuroimaging and Data Science (TReNDS),\\[-0.05cm]
\small  Georgia State University, Georgia Institute of Technology, and Emory University, Atlanta, GA, USA
 \vspace{-0.3cm}
}

\ninept

\begin{document}
\maketitle
\begin{abstract}
Discovering components that are shared in multiple datasets, next to dataset-specific features, has great potential for studying the relationships between different subjects or tasks in functional Magnetic Resonance Imaging (fMRI) data.
Coupled matrix and tensor factorization approaches have been useful for flexible data fusion, or decomposition to extract 
features that can be used in multiple ways.
However, existing methods do not directly recover shared and dataset-specific components, which requires post-processing steps involving additional hyperparameter selection.
\cred{In this paper, we propose a tensor-based framework for multi-task fMRI data fusion, using a partially constrained canonical polyadic (CP) decomposition model. Differently from previous approaches, the proposed method directly recovers shared and dataset-specific components, leading to results that are directly interpretable. A strategy to select a highly reproducible solution to the decomposition is also proposed.}
We evaluate the proposed methodology on real fMRI data of three tasks, and show that the proposed method finds meaningful components that clearly identify group differences between patients with schizophrenia and healthy controls.

\end{abstract}


\section{Introduction}
\label{sec:intro}

Discovering common and distinct features across multiple datasets is a fundamental problem in various disciplines, including the analysis of multi-task/multi-subject fMRI data~\cite{akhonda2021disjointSubspacesCommonDistinctfMRI} or multimodal image fusion~\cite{borsoi2021coupledTensorDecInterImVar}.
The distinct features in each dataset may be generated by variability caused by uncontrolled acquisition conditions~\cite{smilde2017commonDistinctComponentsFusion}, or by features unique to each dataset in, e.g., medical data~\cite{finn2015functionalNetworksFingerprinting}. Accounting for such flexible scenarios is a subject of high interest in data fusion.

Tensor decomposition is of particular interest for data fusion due to its interpretability and strong theoretical foundation. 
Multiple datasets, in many cases, can naturally be represented in the form of high-order tensors, which allow one to take advantage of high-order decomposition models, such as the CP or the PARAFAC2 decomposition~\cite{kolda2009tensor}. 
Coupled tensor decompositions perform data fusion by linking factors among different datasets, benefiting from the complementary information across modalities~\cite{sorber2015structuredDataFusionReview}. Although initial work imposed ``hard couplings'' between multiple modalities~\cite{sorensen2015coupledCPD_part1_unique}, recent work introduced more flexibility by, e.g., weighting the contribution of the coupled components to each dataset~\cite{acar2014structureRevealingFusion}, un-coupling a subset of the columns of one of the factor matrices~\cite{sorensen2020factorizationSharesUnshared,prevost2022LL1_HSR}, or by coupling the factors by constraining them to belong to a Euclidean ball~\cite{farias2016tensorFusionFlexibleCouplings}.
\cred{However, existing coupled decomposition approaches do not consider distinct components (i.e., components that only contribute to a single dataset). This makes them unable to effectively decompose heterogeneous data, in which some components are not shared among the datasets. In this paper, we will address this limitation by proposing a new partially constrained coupled CP decomposition model that directly recovers shared and dataset-specific components.}

In neuroimaging, fusing data from multiple subjects or tasks help elucidate differences in populations and identify putative biomarkers of multiple disorders of the brain, and to understand the brain function in general. 
Previous work investigated the fusion of multisubject data to identify brain patterns common and distinct to different subgroups of subjects for understanding disorders such as schizophrenia~\cite{long2020IVA_commonSubspaceSubgroupsSchizophrenia}.
This has been performed in~\cite{yang2022subgroupIdentification_IVA} using independent vector analysis (IVA)~\cite{kim2006IVA_original}, a multiset extension of independent component analysis (ICA), where the relationship between the different groups of subjects is revealed by the covariance matrix of their cross subject components. 
Both IVA and the PARAFAC2 tensor decomposition have also been recently applied to the analysis of multisubject multi-task fMRI data to reveal components that show differences between patients with schizophrenia and healthy controls, as well as the relationship across multiple tasks~\cite{lehmann2022multiTask_fMRI_PARAFAC2}.
In~\cite{akhonda2021disjointSubspacesCommonDistinctfMRI}, IVA was  used to first estimate the common and distinct subspaces in multi-task fMRI data, and in a second step, the distinct subspaces were analyzed separately using another decomposition, e.g., joint ICA and individual ICA. In~\cite{jin2020dictionaryLearningFMRI_commonIndividual}, dictionary learning was introduced to recover common and distinct components.

In this paper, we propose a tensor-based framework for decomposing multi-task fMRI data into sets of shared and task-specific components. The proposed method integrates the higher-order structure of the data due to the relationship of multiple contrasts used to represent each task with an explicit decoupling between the shared and task-specific components.
Each set of fMRI feature maps corresponding to a single task is ordered in the form of an order-3 tensor, %
which is assumed to follow a low-rank CP model. 
An optimization algorithm is proposed to compute the decomposition. We also propose a new strategy for selecting a highly reproducible result and reducing the dependency of the solution on the initialization of the algorithm~\cite{adali2022reproducibilityReview_SPM}. \cred{Differently from data fusion frameworks that use IVA, which requires
post-processing steps following an IVA decomposition involving additional hyperparameter selection to recover shared and distinct components~\cite{akhonda2021completeIdentificationIVA_fMRIfusion}, the proposed method yields common and shared interpretable components in a single step.}
The proposed methodology is evaluated by decomposing real fMRI data of three different tasks: the Auditory Oddball (AOD), and the Encoding (E) and Probe (P) phases of the Sternberg Item Recognition Paradigm (SIRP)~\cite{gollub2013mcic_fMRI_Dataset}.\footnote{In the rest of the paper, we refer to the E and P phases of SIRP as different tasks since they involve significant differences in data collection.} It is shown that the proposed method finds components that are interpretable, in the sense that the shared and distinct components correspond to activated brain areas in the regions that are expected to be common and distinct across tasks.
This included, for instance, high activations in auditory regions only being present in the AOD task. On the other hand, activations in the default mode network (DMN) region were found to be common among all tasks.
Moreover, various subject factors showed clear group differences between patients with schizophrenia and healthy controls, \cred{along with  spatial maps that are meaningful for the associated tasks.}

\section{Background and Related Work}
\label{sec:related}

\paragraph*{Notation:}

In this work we follow the same general notation and definitions as in~\cite{kolda2009tensor}. Scalars are denoted by lowercase ($x$) or uppercase ($X$) plain font, and vectors, matrices and tensors by lowercase ($\bx$), uppercase ($\bX$) and calligraphic ($\tensor{X}$)  bold font, respectively. %
An order-3 tensor $\tensor{X}\in\amsmathbb{R}^{I\times J\times K}$ is an $I\times J\times K$ array, whose $(i,j,k)$-th element is indexed by $[\tensor{X}]_{i,j,k}$. The $k$-th frontal slice of a tensor $\tensor{X}$ is a matrix whose elements are obtained by fixing its third mode at index $k$, and is denoted by $[\tensor{X}]_{:,:,k}$. The subset of columns of matrix $\bX$ indexed between $i$ and $j$ is denoted by $[\bX]_{:,i:j}$.
The CP decomposition (CPD) of a tensor $\tensor{X}$ with factor matrices $\bA=[\ba_1,\ldots,\ba_R]$, $\bB=[\bb_1,\ldots,\bb_R]$ and $\bC=[\bc_1,\ldots,\bc_R]$ is denoted by
$\tensor{X}=\ldbrack\bA,\bB,\bC\rdbrack = \!\sum_{r=1}^R \ba_r \circ \bb_r \circ \bc_r$, where $\circ$ is the outer product
and $R$ \mbox{is the rank of the CP model.
$\|.\|_F$ is the Frobenius norm.} %

\paragraph*{Tensor methods applied to fMRI:}

In~\cite{chatzichristos2019fMRI_unmixing_highOrderTensor}, the authors consider blind source separation of fMRI data by using an extension of the block term tensor decomposition, where the spatial, temporal, and subject dimensions of the fMRI data are considered as modes of the tensor. However, this uses the raw temporal data and does not benefit from the knowledge of the time regressors, which can be exploited in task-related fMRI data.

Multi-task fMRI data has been analyzed in~\cite{lehmann2022multiTask_fMRI_PARAFAC2} using both IVA and the PARAFAC2 tensor decomposition, both of which were shown to be useful for identifying subgroups of subjects and relationships between different tasks.
Although PARAFAC2 does not make statistical assumptions on the latent components, in contrast to IVA, it imposes additional algebraic conditions on the factors of the mixing model and has different uniqueness conditions.

In~\cite{acar2019biomarkersStructureRevealingTensorFusion}, matrix and tensor decompositions were coupled in the subject mode to fuse EEG, fMRI and sMRI data, where a sparsity regularization in the coefficients modulated the strength of the coupling between different modalities to provide robustness in the presence of modality-specific components. Coupled formulations allow matrix-based approaches to benefit from milder uniqueness conditions of tensor decompositions, providing more interpretable solutions~\cite{sorber2015structuredDataFusionReview,acar2014structureRevealingFusion}.
The uniqueness of coupled matrix decomposition with shared and distinct components was recently studied in~\cite{sorensen2020factorizationSharesUnshared}.

\section{Proposed Method}
\label{sec:proposed}

In this section, we describe the proposed tensor-based strategy for 
shared and distinct multitask fMRI data decomposition. 
We consider $K$ different tasks (e.g., AOD, SIRP, etc.), performed by $S$ subjects. For each task and subject, fMRI data (with $V$ voxels) is acquired in $T_k$ different ways. 
Each one-dimensional feature map is extracted via voxel-wise linear regression of the time-series fMRI data, as in~\cite{levin2017quantifyingInteractionMultipleDatasetsSchizophrenia}.
Because the data of the $T_k$ feature maps for the $k$-th task is expected to contain activations of functional networks in similar brain regions, it is stored in one tensor, or \textit{task dataset}, $\tensor{Y}_k\in\amsmathbb{R}^{S\times V\times T_{k}}$, for $k=1,\ldots,K$.

\paragraph*{Model:}

Our objective is to decompose the fMRI feature maps in $\tensor{Y}_k$ into a set of shared and distinct factors, according to the model
\begin{align}
    \tensor{Y}_k = \tensor{P}_k + \tensor{D}_k \,,
    \label{eq:meas_model}
\end{align}
where tensor $\tensor{P}_k\in\amsmathbb{R}^{S\times V\times T_{k}}$ denotes a component that is partially shared across all task datasets, while $\tensor{D}_k$ denotes a distinct, task-specific component. Each of these tensors is supposed to follow a CP model, where each partially shared component $\tensor{P}_k$ has rank $R$, and the distinct components $\tensor{D}_k$ have rank $L_k$, for $k=1,\ldots,K$. This allows us to write:
\begin{align} \label{eq:cp_model_both}
    \tensor{P}_k = \big\ldbrack \bS^p, \bV^p, \bT_{k}^p \big\rdbrack \,,
    \qquad
    \tensor{D}_k = \big\ldbrack \bS_{k}^d, \bV_{k}^d, \bT_{k}^d \big\rdbrack \,,
\end{align}
where $\bS^p\in\amsmathbb{R}^{S\times R}$ and $\bV^p\in\amsmathbb{R}^{V\times R}$ are the factor matrices of $\tensor{P}_k$ related to the subjects and the voxels (spatial maps), which are shared among all $K$ task datasets, while $\bT_{k}^p\in\amsmathbb{R}^{T_{k}\times R}$ are factor matrices related to the acquisitions of each task, and are thus not shared as the different tasks do not have a direct correspondence. 
The factor matrices $\bS_{k}^d\in\amsmathbb{R}^{S\times L_{k}}$, $\bV_{k}^d\in\amsmathbb{R}^{V\times L_{k}}$, $\bT_{k}^d\in\amsmathbb{R}^{T_{k}\times L_{k}}$ are specific for each task.
The CP model has strong uniqueness properties, which make the factor matrices directly interpretable.

The model proposed in~\eqref{eq:meas_model} and~\eqref{eq:cp_model_both} contains several constraints, which, despite reducing flexibility, aid in the interpretation of the results.
First, there is an explicit separation of partially shared and distinct components across different tensors (datasets): the factors in the first and second mode of $\tensor{P}_k$ are shared among all $K$ task datasets, and the third factor $\bT_{k}^p$ shows the contribution of the shared components to the $k$-th task dataset.
Second, for all $T_k$ fMRI feature maps in the $k$-th task dataset, both the shared and distinct component tensors $\tensor{P}_k$ and $\tensor{D}_k$ share the same spatial maps ($\bV^p$ and $\bV_k^d$) and the same subject factors ($\bS^p$ and $\bS_k^d$), and only their contribution to each fMRI acquisition in the $k$-th task dataset ($[\tensor{Y}_k]_{:,:,i}$, $i=1,\ldots,T_k$) is modulated by the elements of the third-mode factors $\bT_{k}^p$ and $\bT_k^d$.
Thus, $\tensor{Y}_k$ can be written as the following CP model:
\begin{align}
    \tensor{Y}_k = \big\ldbrack \bS_{k}, \bV_{k}, \bT_{k} \big\rdbrack \,,
    \label{eq:cp_yk}
\end{align}
where $\bS_{k}$, $\bV_k$ and $\bT_k$ are related to the factors in~\eqref{eq:cp_model_both} through:
\begin{align}
    \bS_{k} ={} \big[\bS^p,\, \bS_{k}^d \big] \,, \,\,\,\,
    \bV_{k} ={} \big[\bV^p,\, \bV_{k}^d \big] \,,
    \,\,\,\,
    \bT_{k} ={} \big[\bT_{k}^p,\, \bT_{k}^d \big] \,.
    \label{eq:opt_prob_cp2_b}
\end{align}
Note that, assuming $T_k\leq S\leq V$, the CP model~\eqref{eq:cp_yk} is generically unique if $R+L_k\leq(T_k+1)(S+1)/16$~\cite{chiantini2012genericIdentifiabilityTensors}.

\paragraph*{Decomposition algorithm:}

Our aim is to recover the shared and distinct factors corresponding to the subject, voxel and task dimensions, given the multi-task datasets $\{\tensor{Y}_k\}_{k=1}^K$.
To this end, we formulate the flexible coupled tensor decomposition as the following optimization problem:
\begin{align} \label{eq:opt_prob_cp2_full}
    & \min_{\Theta} \,\, J(\Theta) \quad \text{subject to \eqref{eq:opt_prob_cp2_b}} \,,
\end{align}
where the parameter space is given by $\Theta=\big\{\bS_{k}, \bV_{k}, \bT_{k}:1\leq k\leq K\big\}$, the constraint \eqref{eq:opt_prob_cp2_b} ensured the first $R$ columns of $\bS_k$ (resp. $\bV_k$) are the shared for all $k$, and the cost function $J(\Theta)$ is given by %
\begin{align}
    J(\Theta) = \sum_{k=1}^K \Big\|\tensor{Y}_k - \big\ldbrack \bS_{k}, \bV_{k}, \bT_{k} \big\rdbrack \Big\|^2_F + \lambda\big\|\bV_{k}^\top \bV_{k}-\bI\big\|^2_F \,.
    \nonumber
\end{align}
The first term of $J(\Theta)$ consists of a data fitting term, while the sec-\goodbreak\noindent 
ond term consists of a regularization that penalizes the coherence between spatial maps within a task dataset to prevent degenerate solutions containing highly dependent spatial maps, which is a known problem in unconstrained real-valued tensor decomposition~\cite{kolda2009tensor}.

To solve problem~\eqref{eq:opt_prob_cp2_full}, we consider a \emph{block coordinate descent} approach, in which the cost function is iteratively minimized with respect to one group of variables at a time ($\bS_{k}, \bV_{k}$ and $\bT_{k}$, $\forall k$), while keeping the remaining ones fixed with the values from the previous iterations. The optimization w.r.t. $\bS_{k}$ and $\bT_{k}$ consists of constrained least squares problems, which can be solved efficiently. The optimization w.r.t. $\bV_{k}$, on the other hand, is a fourth-order problem; thus, we find a local solution using a quasi-Newton method.

\paragraph*{Best run selection:}

Since problem~\eqref{eq:opt_prob_cp2_full} is non-convex, the solution to the block coordinate descent optimization strategy will depend on the initialization of the algorithm. This makes it more challenging to guarantee that the results are \emph{reproducible}, that is, given the same data and code, we should be able to obtain consistent results~\cite{adali2022reproducibilityReview_SPM}. Reproducibility of the results is very important for their adequate interpretability, particularly in medical imaging.
In this work, inspired by~\cite{adali2022reproducibilityReview_SPM}, we consider a heuristic strategy to select the ``most reproducible'' run from a set of solutions to problem~\eqref{eq:opt_prob_cp2_full} obtained from different random, independent initializations.
First, we solve~\eqref{eq:opt_prob_cp2_full} using $N$ random initializations, and store each solution in a set $\Omega$. Then, we compute a similarity (a pseudo-metric) between each pair of solutions $\{\bS_k,\bV_k,\bT_k\}$ and $\{\bS_k',\bV_k',\bT_k'\}$ in $\Omega$ as:
\begin{align}
    & 
    {\scriptsize {\rm pdistance} = -\sum_{k=1}^K \min_{\sigma\in\Sigma_{R+L_k}} \bigg\{ \sum_{r=1}^{R+L_k} \bigg( 
    \frac{[{\bS}_k]_{:,r}^\top
    [{\bS}_k']_{:,\sigma(r)}}{\|{\bS}_k\|\|{\bS}_k'\|} 
    }
    \nonumber \\ & \qquad\qquad
    {\scriptsize 
    + \frac{[{\bV}_k]_{:,r}^\top
    [{\bV}_k']_{:,\sigma(r)}}{\|{\bV}_k\|\|{\bV}_k'\|} 
    + \frac{[{\bT}_k]_{:,r}^\top
    [{\bT}_k']_{:,\sigma(r)}}{\|{\bT}_k\|\|{\bT}_k'\|} 
    \bigg) \bigg\} 
    }
    \,,
    \label{eq:pseudometric_reproducib}
\end{align}
where $\Sigma_{R+L_k}$ is the set of all permutations of $\{1,\ldots,R+L_k\}$. The minimization is performed as a linear assignment problem~\cite{duff2001algorithmsAssignementProblemSparse}.
Finally, we select the most reproducible solution as the one that is the most similar to every other solution in $\Omega$, according to the similarity criterion defined in~\eqref{eq:pseudometric_reproducib}.
The procedure is described in Algorithm~\ref{alg:alg_cp}.

\section{Experiments}
\label{sec:experims}

\paragraph*{Dataset:}

We considered two fMRI datasets from the MCIC collection~\cite{gollub2013mcic_fMRI_Dataset}, which are collected from 271 subjects (121 being patients suffering from schizophrenia (SZ) and 150 being healthy controls (HC)) that perform an AOD and SIRP tasks.
From AOD and SIRP datasets, three task datasets were generated as follows.
For each subject and task, lower-dimensional features were extracted from the raw fMRI data by using regressors obtained by convolving the hemodynamic response function (HRF) in the SPM toolbox~\cite{SPM12software} with the desired predictors, as in~\cite{levin2017quantifyingInteractionMultipleDatasetsSchizophrenia}. The $T_k$ regression coefficient maps for the $k$-th task were then ordered as tensor $\tensor{Y}_k$, containing $S=271$ subjects, $V=48546$ voxels and $T_k$ feature maps.
In the AOD task, each subject listened to three different types of stimuli (standard, novel and target), which were randomly ordered, and was required to press a button when the ``target'' stimulus occurred. Three fMRI feature maps were extracted for this task by using regressors corresponding to the target [T], novel [N], and target with the standard [TS] stimuli, resulting in the frontal slices of $\tensor{Y}_1\in\amsmathbb{R}^{271\times48546\times3}$ ($T_1=3$).
The SIRP task was divided into two different task datasets, corresponding to the Encode (SIRP-E) and to the Probe (SIRP-P) phases of the experiment, respectively. In the encoding phase, subjects needed to memorize a set of integer digits between 0 to 9, and in the probe phase, the subjects were shown a sequence of digits and were asked to press a button when a digit belonged to the memorized set. This experiment was performed with 1, 3 and 5 digits in the set, resulting in three distinct fMRI feature maps for each phase, which are ordered as the frontal slices of $\tensor{Y}_2\in\amsmathbb{R}^{271\times48546\times3}$ (for SIRP-E, with $T_2=3$) and $\tensor{Y}_3\in\amsmathbb{R}^{271\times48546\times3}$ (for SIRP-P, with $T_3=3$).

\begin{algorithm} [t]
\footnotesize
\SetKwInOut{Input}{Input}
\SetKwInOut{Return}{Return}
\caption{Coupled CP-based fMRI fusion\label{alg:alg_cp}}
\Input{$\{\tensor{Y}_k\}_{k=1}^K$, ranks $R,\{L_{k}\}_{k=1}^K$, $\lambda$, $\Omega=\varnothing$, $N$.}

\For{$n=1,2,\ldots,N$}{
Initialize ${\bS}_k^{(0)}$, ${\bV}_k^{(0)}$, ${\bT}_k^{(0)}$ randomly, set $i=0$ \;

\While{Stopping criterion is not satisfied}{
$i\leftarrow i+1$ \;

${\bS}_k^{(i)}\leftarrow$ Minimize $J(\Theta)$ s.t.~\eqref{eq:opt_prob_cp2_b} w.r.t. $\{{\bS}_k\}_{k=1}^K$, with ${\bV}_k={\bV}_k^{(i-1)}$ and ${\bT}_k={\bT}_k^{(i-1)}$\;

${\bV}_k^{(i)}\leftarrow$ Minimize $J(\Theta)$ s.t.~\eqref{eq:opt_prob_cp2_b} w.r.t. $\{{\bV}_k\}_{k=1}^K$, with ${\bS}_k={\bS}_k^{(i)}$ and ${\bT}_k={\bT}_k^{(i-1)}$\;

${\bT}_k^{(i)}\leftarrow$ Minimize $J(\Theta)$ w.r.t. $\{{\bT}_k\}_{k=1}^K$, with ${\bS}_k={\bS}_k^{(i)}$ and ${\bV}_k={\bV}_k^{(i)}$\;
}
$\Omega\leftarrow \Omega \bigcup \big\{{\bS}_k^{(i)},{\bV}_k^{(i)},{\bT}_k^{(i)}\big\}$ \;
}

\mbox{${\bS}_{k}, {\bV}_{k}, {\bT}_{k} \leftarrow$ Pick the most reproducible run in $\Omega$ according to~\eqref{eq:pseudometric_reproducib}\;}

\Return{${\bS}_{k}, {\bV}_{k}, {\bT}_{k}$, $k=1,\ldots,K$.}

\end{algorithm}

\begin{figure*}%
    \includegraphics[width=1\linewidth]{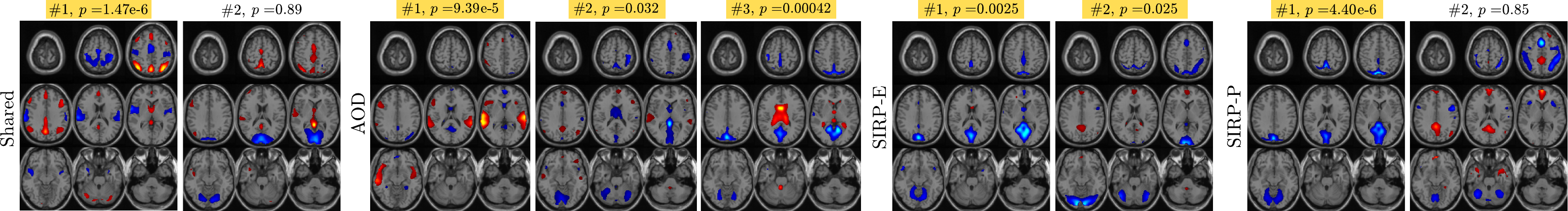} \\
    \vspace{-0.73cm}
    \caption{Estimated shared and distinct components resulting from the fusion of the AOD, SIRP-E and SIRP-P task datasets.
    The shared components show high activations in the DMN, visual, motor and frontoparietal regions. In the distinct components corresponding to the three task datasets, we observe: \textit{i)} high auditory and visual activations in the AOD task; \textit{ii)} high visual activations in the SIRP-E task; \textit{iii)} high visual, frontoparietal, motor and DMN activations in the SIRP-P task.}
    \label{fig:fus_AOD_SIRPE_SIRPP}
    \vspace{-0.08cm}
\end{figure*}
\begin{figure*}%
    \centering
    \includegraphics[width=0.775\linewidth]{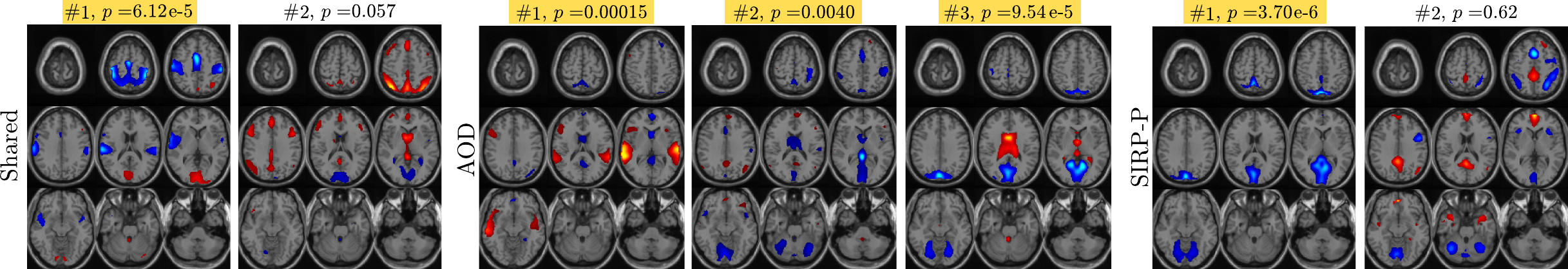}\\
    \vspace{-0.35cm}
    \caption{Estimated shared and distinct components resulting from the fusion of the AOD and SIRP-P task datasets.
    The shared components show high activations in the motor, DMN and frontoparietal regions. In the distinct components corresponding to the two task datasets, we observe: \textit{i)} high auditory and visual activations in the AOD task; \textit{ii)} high visual, frontoparietal, motor and DMN \mbox{activations in the SIRP-P task.}}
    \label{fig:fus_AOD_SIRPP}
    \vspace{-0.08cm}
\end{figure*}
\begin{figure*}[!t]
    \centering
    \includegraphics[width=0.67\linewidth]{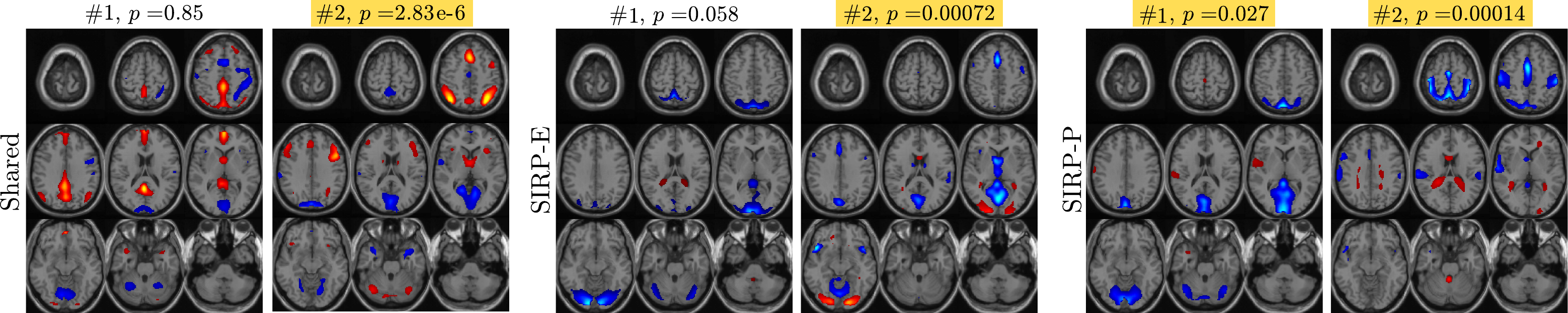}\\
    \vspace{-0.35cm}
    \caption{Estimated shared and distinct components resulting from the fusion of the SIRP-E and SIRP-P task datasets.
    The shared components show high activations in the DMN, angular gyrus and frontoparietal regions. In the distinct components corresponding to the two task datasets, we observe: \textit{i)} high visual activations in the SIRP-E task; \textit{ii)} high visual and motor activations in the SIRP-P task.}
    \label{fig:fus_SIRPE_SIRPP}
    \vspace{-0.05cm}
\end{figure*}

\paragraph*{Method setup:}

We selected the ranks of the decomposition so as to obtain the 
best reproducibility according to the criterion~\eqref{eq:pseudometric_reproducib}. This resulted in $R=2$ for the shared component, and $L_1=5$ (AOD), $L_2=4$ (SIRP-E) and $L_3=4$ (SIRP-P) for the distinct components.
The regularization parameter was selected similarly as $\lambda=10^{6}$. Algorithm~\ref{alg:alg_cp} was implemented in Matlab\textsuperscript{TM} and executed in a computer with four 3.2Ghz cores and 24Gb RAM. $N=200$ reproducibility runs were used. Before applying Algorithm~\ref{alg:alg_cp}, the subject and voxel modes of tensors $\tensor{Y}_k$ are compressed to a dimension of $30$ using the SVD in order to reduce its complexity~\cite{bro1998cpd_svd_compression}. 
The total execution time for all experiments was 157.5 seconds.

\paragraph*{Results:}

The components (normalized to unit variance and thresholded at $|z|=2.7$) obtained by fusing the three task datasets are shown in Fig.~\ref{fig:fus_AOD_SIRPE_SIRPP}, and the results of only fusing AOD and SIRP-P, or SIRP-E and SIRP-P, are shown in Figs.~\ref{fig:fus_AOD_SIRPP} and~\ref{fig:fus_SIRPE_SIRPP}, respectively. A two sample t-test on the subject factors was used to evaluate whether significant ($p<0.05$) group differences between HCs and SZs are present.
Red/yellow voxels indicate a higher activation in controls than in patients (determined according to the result of the t-test), while blue voxels mean the opposite.

\textit{AOD, SIRP-E and SIRP-P (Fig.~\ref{fig:fus_AOD_SIRPE_SIRPP}):} Shared component \#1 contains activations in the DMN, motor and frontoparietal (FP) regions, which shows highly significant differences between SZs and HCs. Shared component \#2 contains activations in the visual region without a significant difference between SZs and HCs.
The distinct components for AOD are significant, and contain activations in auditory (\#1), motor (\#2), and visual (\#3) regions.
The components \#1 and \#2 of SIRP-E contain high activations in the visual regions. While component \#1 of SIRP-P also contains visual activations, component \#2 of SIRP-P contains high FP, motor and DMN activations that do not appear as strongly in the components of SIRP-E.

\textit{AOD and SIRP-P (Fig.~\ref{fig:fus_AOD_SIRPP}):} Shared component \#1 is significant and contains high activations in the motor area, whereas shared component \#2 contains high activations in the DMN and the FP regions but does not show a significant difference between SZs and HCs.
The distinct components of both  AOD and SIRP-P  are  similar to those obtained with all three datasets with differences in $p$-values. %

\textit{SIRP-E and SIRP-P (Fig.~\ref{fig:fus_SIRPE_SIRPP}):} Shared component \#1 contains activations in the DMN and angular gyrus (AG) regions, and is not significant. Shared component \#2 is highly significant, with a high activation in the FP region.
The distinct components of SIRP-E have high activation in the visual region, but only component \#2 shows a significant difference between SZs and HCs. 
Both components of SIRP-P are significant, with components \#1 and \#2 containing high activations in the visual and motor regions, respectively.

\section{Discussion}

The shared components obtained by the proposed method contain high activations in the DMN, FP and AG regions. 
These activations are highly similar across the shared components obtained from the three sets of fusion results.
Motor responses appear in a shared component when both AOD and SIRP-P are among the tasks being fused, and in a distinct component when only SIRP-E and SIRP-P are fused. This is a nice confirmation for the methodology because the AOD and SIRP-P tasks contain a motor component, which is not present in SIRP-E. 
\cred{From Figs.~\ref{fig:fus_AOD_SIRPE_SIRPP},~\ref{fig:fus_AOD_SIRPP} and~\ref{fig:fus_SIRPE_SIRPP}, the motor component shows significant group difference ($p<0.05$) between HCs and SZ patients. However, when the motor component is estimated as shared (Fig.~\ref{fig:fus_AOD_SIRPP}), it shows a more significant difference between HCs and SZs. This indicates that, when fused, the datasets with motor movements (AOD and SIRP-P) influence each other to estimate a motor component with higher group difference level, compared to when they are fused with a dataset that does not contain motor movements (SIRP-E). This shows the advantage of fusing datasets which are more similar. Moreover, the $p$-values of the motor components easily stay significant even after conservative corrections for  multiple comparisons, like the Bonferroni method.}
Auditory components are only present in the distinct components of AOD, and are highly significant. \cred{AOD is frequently used to differentiate between HCs and SZs, and here it also provides the lowest $p$-values and most discriminative components, with differences between subjects and patients in the areas that are expected.} Visual activations are observed in distinct components for all task datasets, which is reasonable since the subjects performed the tasks with their eyes open.

The activations in various distinct components in Fig.~\ref{fig:fus_AOD_SIRPE_SIRPP} (\#1 and \#2 for AOD, \#2 for SIRP-P), are similar to those obtained by fusion of the same dataset using IVA and PARAFAC2 reported in~\cite{lehmann2022multiTask_fMRI_PARAFAC2}, although their p-values are slightly lower than those of the components that were recovered by IVA.
Hence, even with differences in modeling assumptions the main conclusions in such decompositions agree. This suggests that matrix and tensor decompositions with minimal assumptions on the relationship among the datasets yield useful and directly interpretable results. In our case, we provide components that also report on shared and distinct aspects of the multiple datasets without having to determine additional steps to identify these. This is in contrast to IVA, where deciding which components are shared and which are distinct has to be performed as a post-processing step, which also involves additional hyperparameter selection~\cite{akhonda2021completeIdentificationIVA_fMRIfusion}.
Note, however, that comparing the results of different decomposition methods is very challenging, since they often need vastly different orders to produce meaningful components. Nonetheless, such comparisons are important for providing guidance on the selection of best method for a given dataset.

\clearpage

\bibliographystyle{IEEEbib}
\bibliography{references_fMRI,references_theoretical}

\end{document}